%Paper: hep-th/9401079
%From: polonyi@crncls.in2p3.fr
%Date: Mon, 17 Jan 94 15:35:12 +0100

\magnification= \magstep 1
%\magnification=1200
%Equation numbering macro
\global\newcount\meqno
\def\eqn#1#2{\xdef#1{(\secsym\the\meqno)}
\global\advance\meqno by1$$#2\eqno#1$$}
\def\pr{{\it Phys. Rev.}}
\def\pre{{\it Phys. Rep.}}
\def\re{{\it Rev. Mod. Phys.}}
\def\np{{\it Nucl. Phys.}}
\def\pl{{\it Phys. Lett.}}
\def\gellow{M. GellMann and F. Low, \pr  {\bf 95} (1954) 1300.}
\def\wilsrg{K. Wilson and J. Kogut, \pre {\bf 12 C} (1974) 75 ;
K. Wilson, \re {\bf 47} (1975) 773. }
\def\zinnj{E. Br\'ezin, J.C. Le Guillow and J. Zinn-Justin,
{\it Field Theoretical Approach to Critical Phenomena} in
{\it Phase Transitions and Critical Phenomena} vol.6,
C. Domb and M.S. Green eds. (Academic Press, New York 1976).}
\def\chiran{S.L. Adler, \pr {\bf 177} (1969) 2426 ;
J.S. Bell and R. Jackiw, {\it Nuovo Cimento} {\bf A 60}(1969) 47.}
\def\mirans{P.I. Fomin, V.P. Gusynin, V.A. Miranski and YU.A. Sytenko,
{\it Riv. Nuovo Cimento}, {\bf 6}, No. 5 (1983).}
\def\bardeen{C.N. Leung, S.T. Love , William A. Bardeen,
\np {\bf B 323} (1989) 493 ;
C.N. Leung, S.T. Love , William A. Bardeen , \np {\bf B 273} (1986) 649.}
\def\thoofi{G. 't Hooft, \pr {\bf D 14} (1976) 3432. }
\def\polyi{A. Polyakov, \np {\bf B 120} (1977) 429. }
\def\paris{J. Polonyi, in the Proceedings of the Workshop on
{\it QCD Vacuum Structure and Its Applications}, H.M. Fried and
B. Muller eds., World Scientific, 1993 pag. 3.}
\def\zinnin{E. Br\'ezin, J.C. Le Guillou and J. Zinn-Justin,
\pr {\bf 14} (1976) 2615.}
\def\reisz{Reisz, \np {\bf B 358} (1989) 417.}
\def\belpol{ A.Belavin and A.Polyakov, JEPT Letters {\bf 22} (1975) 245. }
\def\jevicki{ A.Jevicki, \np {\bf B127} (1977) 125.}
\def\polyjept{ A.Polyakov, JEPT {\bf 41} (1975) 988. }
\def\polypl{ A.Polyakov, \pl {\bf B59} (1975) 79. }
\def\forster{ D.Forster, \np {\bf B130} (1977) 38.}
%
% Reference numbering macro
\global\newcount\refno
\def\ref#1{\xdef#1{[\the\refno]}
\global\advance\refno by1#1}
\global\refno = 1
\hoffset =0.8 true cm
\hsize = 15.8 true cm
\vsize = 21 true cm
\tolerance 10000

%
% vector=boldface

\def\ln{{\rm ln}}
\baselineskip=0.1cm
\baselineskip 12pt plus 1pt minus 1pt
\noindent LPT 93-3\hfill September 1993
\vskip .5in
\centerline{\bf MINI-INSTANTONS}
\vskip 24pt
\centerline{Vincenzo Branchina
\footnote{$^a$}{
On leave from Catania University, and INFN Sezione di Catania, Italy}
and
Janos Polonyi
\footnote{$^b$}{
On leave from L. E\"otv\"os University and CRIP, Budapest, Hungary}}
\vskip 12pt
\centerline{\it Laboratory of Theoretical Physics}
\centerline{\it Department of Physics}
\centerline{\it Louis Pasteur University}
\centerline{\it 67087\ \ Strasbourg\ \ Cedex\ \ France}
\centerline{\it and}
\centerline{\it CRN\ \ Strasbourg }
\centerline{\it 67037\ \ Strasbourg\ \ Cedex\ \ France}
\vskip 1.5in
%\vfill
\centerline{Submitted to {\it Nuclear Physics B\/}}
%\bigskip
%\eject

%\baselineskip 24pt plus 2pt minus 2pt
\vskip 24 pt
\baselineskip 12pt plus 2pt minus 2pt
\centerline{{\bf ABSTRACT}}
It is shown that the inhomogeneous saddle points of scale invariant
theories make the semiclassical expansion sensitive on the choice of
non-renormalizable operators. In particular, the instanton
fugacity and the beta function of the two dimensional non-linear sigma model
depends on apparently non-renormalizable operators. This represents a
non-perturbative breakdown of that concept of
universality which is based on low dimensional
operators.
\medskip
\vfill
\eject
\noindent{\bf I. INTRODUCTION}
\medskip
\nobreak
\xdef\secsym{1.}\global\meqno = 1
\medskip
The ultraviolet divergences of Quantum Field Theories is an evergreen
subject for new discoveries and frustration. On the one hand,
it motivated the introduction of the renormalization \ref\gellowr\
\ref\wilsrgr\ which subsequently proved to be an essential
notion, independently of the divergences. It was used to uncover how
effects due to different interactions build up as the function of their
length scale. This kind of thinking
led to the application of universality in field theory
\wilsrgr, \ref\zinnjr. It states that
it is enough to consider renormalizable models because the non-renromalizable
coupling constants decrease as we move towards the infrared direction
and become negligible in the regime where the measurements are performed.
On the other hand, the original naive hope that
the physics at the ultraviolet cut-off decouples from the observations
as the cut-off is removed turned out out be unjustified by the
isolation of the anomalies \ref\chiranr. The somewhat unfortunate
name for the survival of the cut-off effects in renormalized models came from
the surprising and apparently unrelated manner this phenomenon arises in
different regularization schemes. After decades of intensive investigations
one is still uncomfortable in this subject since the true physical
origin and the necessity of the anomalies has not yet been understood.
But at least the class of models where the anomalies respect renormalizability
and universality has been established by means of involved algebraic and
topological methods.

Another observation concerning the importance of the cut-off effects
has been made for non-asymptotically free models. If the coupling
constant is large at the cut-off then non-perturbative phenomena
may generate novel infrared behavior. Consider say QED as the
cut-off is removed \ref\miransr.
This is certainly a non-perturbative problem due to
the Landau-pole and we are left for speculations only. Suppose that
the coupling constant of massless QED is so
large at short distances that the electron-positron pair is bound
with negative energy. These small bound states then condense in the
vacuum and lead to the dynamical breakdown of the chiral symmetry.
The loss of the chiral symmetry is obviously a long distance phenomenon.
Thus the bound state formation at the cut-off influences the infrared
behavior of the model. The methods to incorporate bound state
effects are not powerful enough so the same phenomenon was later established
by means of the renormalization group \ref\bardeenr.
The description of the role of
an operator in the renormalization process starts with the power counting
arguments. It was suggested that we should include the anomalous dimension
in these considerations since from the point of view of the scaling behavior
there is no difference between the canonical and the anomalous dimensions.
The power coupling argument applied with anomalous dimensions which were
computed in the weak coupling expansion and became extended to the strong
coupling regime brought a surprise: the operators
which control the dynamical breakdown of the chiral symmetry appear
renormalizable. The pile-up of the graphs in the ultraviolet regime
observed in the solution of the renormalization group generates those coupling
constants which are needed for the chiral symmetry broken vacuum.
The net result is that some new operators turn out to be relevant in
describing the physical content of the theory.

We present another cut-off phenomenon which generates new important
coupling constants in addition to the perturbatively renormalizable
ones. The novel feature of this phenomenon is its simplicity and
that it occurs in asymptotically free theories. It is believed that the
smallness of the bare coupling constant in asymptotically free theories
renders all cut-off effects perturbative. It will be shown below that
this expectation is not always fulfilled. We found new operators
mixed in the renormalization of such scale invariant theories which possess
nontrivial saddle points.

As an example consider the saddle point approximation of QCD.
The conventional approach is based on the renormalized perturbation
expansion \ref\thoofir, \ref\polyir. One first isolates the extrema of
the renormalized action,
\eqn\sren{S_r[A]={1\over4g^2_r}\int d^4xF^2_{\mu\nu}
={1\over4g^2_r}\int d^4xL[A],}
the instantons. In the next step
the one-loop quantum correction is computed. It is ultraviolet divergent
so counterterms $S_{ct}[A]$ are needed. Fortunately it turns out that
for instantons which are locally flat the counterterms are
the same as in the zero winding number sector so the renormalization
procedure remains unchanged .

We should not forget that this procedure is rather formal, the application of
the renormalized perturbation expansion is usually difficult to justify.
The mathematically solid framework is that of the bare cut-off theory
described by the well defined path integral,
\eqn\bpt{\int D[A]e^{-S_b[A]}.}
which includes the bare action,
\eqn\bra{S_b={1\over4g^2_b}\int d^4xL[A]=S_r+S_{ct}.}
\bpt\ is an integral with large but finite dimension. We are interested
in the limit when its dimension diverges in order to arrive the renormalized
theory. The bare perturbation expansion is obtained by expanding
\bpt\ in the non-quadratic parts of $S_b[A]=S_0[A]+S_1[A]$,
\eqn\bptk{\int D[A]e^{-S_0[A]+S_1[A]}
=\int D[A]e^{-S_0[A]}\sum{1\over n!}S^n_1[A].}
The small parameter of the expansion is the
bare coupling constant, $g_b$. As the cut-off is removed the bare parameters
must be rescaled in order to keep the physical content of the theory cut-off
independent. After the expansion one recognizes some cancellation between
different contribution in the sum of \bptk. After taking these cancellations
into account we arrive to a new expression, the renormalized perturbation
expansion which looks as if it had come
from the expansion of $S_r[A]$ according to a small parameter of the
renormalized action, $g_r$ the renormalized coupling constant.

But this
is not quite right. The more precise statement is that one first expands
$S_r[A]$ and then retains only the `finite' parts of the graphs. The
`divergent' pieces are canceled by the counterterms. The problem is that
there is no such action which would generate only the finite contribution
in the perturbation expansion. In order to prove the
existence of the renormalized perturbation expansion one first has to
establish the reliability of the bare perturbation expansion. It is
only after this latter step that one can take the left over
'renormalized' part of the perturbation expansion seriously.

One would think that the renormalized perturbation expansion
is reliable in QCD. In fact, the one-loop relation
\eqn\rbrc{{1\over g^2_b}={1\over g^2_r}-2\beta_0\ln{\mu\over\Lambda},}
where $\mu$ and $\Lambda$ are the subtraction scale and the cut-off,
respectively, indicates that the counterterms are suppressed compared to
the renormalized action,
\eqn\abrk{S_b[A]={1\over g^2_r}\int d^4xL[A]
-2\beta_0\ln{\mu^2\over\Lambda}\int d^4xL[A],}
for the sufficiently large region
\eqn\sreg{{\mu\over\Lambda}<e^{1\over2\beta_0 g^2_r}.}

Alas, there is a problem with this argument in the nonzero
winding number sector. According to the renormalized perturbation expansion
the tree level step, the selection of the saddle points of
\eqn\sbrp{\int D[A]e^{-S_r[A]+S_{ct}[A]}}
is done by the help of the renormalized part, $S_r[A]$ and
the counterterms are included perturbatively only at the one-loop level.
Since the theory is scale invariant the renormalized action has no
scale parameter and the saddle points form a degenerate family with
respect to the scale transformation. The scale invariance is broken by
the quantum corrections, the well known phenomenon of the dimensional
transmutation. In the contrary, the scale invariance is broken
already at the tree level in the bare theory. This seems rather
unusual in the dimensional regularization scheme
where the usually employed minimal subtraction hides the cut-off
in the computations. But the less formal, non-perturbative
regularization schemes reveal the scale dependence of the bare
theory.

Consider first the regularization by adding higher order
derivatives to the action,
\eqn\hdrf{S_b[A]=\int d^4xF_{\mu\nu}
\biggl\{{1\over4g^2_b}+{C_2\over\Lambda^2}D^2
+{C_4\over\Lambda^4}D^4+\cdots\biggr\}F_{\mu\nu},}
where $D$ stands for the covariant derivative. Though these higher dimensional
operators do not remove all the UV divergences from the theory they
serve as an example for our purpose. The new terms play no role
in the physical content of the theory according to the perturbation
expansion. This is because each insertion of the new vertices bring the
coefficient
\eqn\insv{C_n\biggl({ \mu\over\Lambda}\biggr)^n,}
where $\mu$ is the characteristic energy scale. The same scale dependence
generates ultraviolet divergences and may make the theory non-renromalizable.
The instanton with scale $\rho$ has the action
\eqn\instahd{S_i=8\pi^2\biggl\{{1\over4g^2_b}
+C_2O\biggl({1\over\rho^2\Lambda^2}\biggr)
+C_4O\biggl({1\over\rho^4\Lambda^4}\biggr)+\cdots\biggr\}.}
It is not surprising that the
scale degeneracy of the instantons is split in the bare action since
after all it contains a scale parameter, the cut-off. The situation is
similar in lattice regularization,
\eqn\instla{S_i={2\pi^2\over g^2_b}+O(\rho^2a^2),}
where the lattice spacing is denoted by $a$. In the systematical
saddle point expansion of the bare theory the scale dependence appears
already at the tree level in a manner radically different from the
one-loop scaling violation of the minimal subtraction scheme.
It is certainly true that the magnitude of the `counterterms',
i.e. the cut-off dependent terms of the action is negligible
compared to the renormalized one but this is not enough now.
It is not enough because the renormalized perturbation expansion
has an infinite degeneracy at the tree level with respect to the
scale transformation. An arbitrarily small perturbation which
lifts this degeneracy may have substantial effect in this situation.
Similar phenomenon can be observed in the quantum Hall effect where
degeneracy of the Landau levels is lifted in a very specific
manner by the electrostatic interaction.

The weak scale dependence
of the instanton action generated by the cut-off is enough to destabilize
several saddle points or make only some of them stable.
When
\eqn\instins{C_2>0,\ \ C_4=C_6\cdots=0}
is chosen then the instantons grow without limit. For
\eqn\stinst{C_2<0,\ \  C_4>0,\ \ C_6=\cdots0,}
there is a stable saddle point, called
mini-instanton with the size parameter $\bar\rho=O(1/\Lambda)$.
On the lattice the instanton action tends to zero for small instantons
which `fall through the lattice'. Thus depending on the choice of the
lattice action we may or may not have stable saddle points. The tree
level instanton action is sensitive to the irrelevant i.e. non-renormalizable
operators present in the action.

Since the regularization is achieved
by the introduction of some specific irrelevant operators, different
regularizations lead to different saddle point structure. For large
saddle points the presence of the cut-off lead to small changes. But
for small saddle points whose size is comparable with the cut-off
the non-renormalizable operators are important. By an appropriate
choice of the non-renormalizable coupling constants the
scale degeneracy of the renormalized action is split in such a manner
that the stable saddle points will be at the scale of the cut-off.
In this case the saddle point, being close to the cut-off, reflects
the physics of short distances. One recognizes the analogy with
strong coupling QED at this point. There the strong attraction at
short distances led to the formation of small bound states. These
bound states populate the vacuum and lead to long distance effects.

In QCD the (weak) interactions at short distances lead to new saddle
points.
Can the modification of the instanton action at short distances
change the physics at large distances ? It is difficult to answer
this question before actually completing the saddle point expansion.
At the first moment one would think that the presence of instantons
with small size only will certainly modify the long range structure.
The disappearance of the degeneracy corresponding to
the dilatation transformation raises even the hope of an
infrared consistent instanton gas description of the vacuum.
But this is not necessarily so. The integration over the dilatation
mode can be performed by the collective coordinate method as usual.
Then for a given value of the instanton size the collective coordinate keeps
the saddle point stable for the Gaussian integration. As we integrate
over all values of the instanton size we should ultimately recover the
contribution of large saddle points. This is the same as in the
renormalized perturbation expansion. We are now confronted with the
question whether the small or the large instantons dominate the integral
over the scale parameter. The action of the small stable instanton
in the case \stinst\ is smaller than the asymptotic value achieved for
large size. Thus the small instanton contribution is enhanced.
We believe that as long as the integration over the scale parameter
is done only in the `safe' region,
\eqn\safere{\rho<<\Lambda_{QCD}^{-1},}
far from the confinement radius the effect of the small instantons
will be felt. The coupling constants generated by them might be
responsible for the restoration of the center symmetry which is
usually broken by the gauge fixing condition and
crucial for the understanding of the long range features of the
vacuum in the framework of the renormalization group \ref\parisr.

One might argue that such a strange dependence of the physics
on certain higher dimensional operators is not important if
one uses the lagrangian which does not contain them from the very beginning.
This is certainly true mathematically but in the real world we always have
irrelevant operators in our theories. Denote the theory we investigate
by $T_1$. Unless it is the Theory of Everything there is always a higher energy
unified theory, $T_2$,
which gradually becomes relevant as the energy of the observations is
increased. Thus the renormalized trajectory does not
approach the ultraviolet fixed point of $T_1$ but rather turns to
another direction, given by $T_2$. The interactions of $T_2$
generate vertices when expressed in terms of the degrees of
freedom of $T_1$. These effective vertices represent non-renormalizable
irrelevant operators in $T_1$ since otherwise the renormalized trajectory
would run into the fixed point of $T_1$. Thus the `seeds' of the new physics
are always hidden in the irrelevant operators of the theories we used
to describe the high energy experiments. Expressing it in mathematical
terms: the theory $T_2$  formally regulates the ultraviolet region of
$T_1$ and the ultraviolet regulators are always irrelevant operators.
Thus we have irrelevant operators in a given regularization scheme even
if they do not appear in the formal, unregulated expression of the lagrangian.

In order to demonstrate the above peculiarities of the semiclassical
expansion we computed the weight of the one-instanton
sector in the two dimensional sigma model.
This model is usually defined by the action
\eqn\sakez{S[\vec\phi]={1\over2f}\int d^2x(\partial_\mu\phi^a)^2,}
where the three component field $\phi^a$, $a=1,2,3$ is subject of the
constraint
\eqn\constr{\sum_{a=1}^3\phi^a\phi^a=1.}
Since there is no other renormalizable operator with the same symmetry
the action \sakez\ is believed to cover the universality class
of the model. It will be shown that this is not what
happens. \sakez\ is usually investigated by means of the weak coupling
expansion. Since there are instanton solutions one has to
include their contributions as well. It turns out that in the nonzero
winding number sector some naively non-renormalizable operators
play role in the evolution of the coupling constant, $f$.

We extend the action \sakez\ by adding higher dimensional terms,
such as $(\partial_\mu\phi^a)^{2n}$, $n=2,3$ to the lagrangian,
\eqn\exta{S[\vec\phi]={1\over2f}\int d^2 x~\Biggl
( 1- {3\over 4}~{C_2\over{\Lambda^2}}(\partial_\mu\phi^a)^2
+ {5\over 64}~{C_4\over{\Lambda^4}}(\partial_\mu\phi^a)^2
(\partial_\mu\phi^a)^2\Biggr)(\partial_\mu\phi^a)^2,}
where $\Lambda$ is the euclidean momentum cut-off. The sign
of the quartic term is fixed in \exta\ by anticipating the choice of the
non-renormalizable operators which actually influence the physical
content of the model. The numerical coefficients were chosen
for later convenience.

Though there are non-renormalizable operators in \exta\ the theory
is still renormalizable. There are no overall divergences generated
by the new pieces. In fact, each additional derivative is compensated
by a factor of $1\over\Lambda$ and the insertion of the new
vertices into a graph does not increase the overall degree of divergence.
The issue of the overlapping divergences is much more complex and one
has to use the O(3) symmetry \ref\zinninr. It can be proven by
induction \ref\reiszr\ that in each order of the loop expansion
the ultraviolet divergences can be eliminated by the fine tuning the
coupling constant $f$ for $any$ action which reduces to \sakez\ classically as
the cut-off is removed. Naturally the constants $C_n$ become nontrivial
functions of the cut-off as one leaves the vicinity of the ultraviolet
fixed point.

It is important to bear in mind that the
proof of the renormalizability holds only for the expansion around a
homogeneous configuration. When the background around which we expand
is inhomogeneous then the situation is more involved. The new vertices
in \exta\ leave the overall divergence unchanged only if the characteristic
length scale $\rho$, of the background field configuration
is far from the cut-off, $\rho\Lambda>>1$. The generalization
of the inductive proof for the overlapping divergences for the
case of inhomogeneous background field is not known and one expects
some problem there, too, when $\rho\Lambda=O(1)$.

Consider a physical quantity, $P$, defined by the help of the scale
parameter $\mu$ with dimension of energy. In case of a Green function
$\mu$ can be taken to be the typical energy scale of the external legs. For
a bulk quantity $\mu^{-1}$ is the characteristic length scale of the
system. The observable depends on the coupling constants and the
cut-off,
\eqn\iratg{P=\Lambda^{[P]}\tilde P(f,C_n\mu^n/\Lambda^n),}
where $[P]$ is the dimension of $P$ in units of energy and $\tilde P$
is a dimensionless function. If the scales of the
observable and the cut-off are far, $\mu/\Lambda\approx0$,  then
dependence of $P$ on the non-renormalizable coupling constants is negligible.
In this paper the ratio
\eqn\irat{P={Z_1\over Z_0},}
will be computed where $Z_n$ denotes the path integral
for the field configurations with winding number $n$. Since \irat\ is a
bulk quantity  its dependence on $C_n$ should be negligible for large enough
value of the ultraviolet cut-off, $\Lambda$,
\eqn\iratp{P=\tilde P(f)=F(\Lambda_\sigma/\Lambda),}
where $\Lambda_\sigma$ is the cut-off independent lambda-parameter of the
model. This is not what happens in our case, the sensitivity of
the ratio \irat\ on $C_n$
survives the removal of the cut-off. For an observable
computed on a homogenous or slowly varying background field the
dependence on the non-renromalizable coupling constant, \insv,
is suppressed. But the ratio \irat\ is saturated by the fluctuations
around a mini-instanton and the cut-off effects survive the renormalization.

Since $P$ must be
finite as the cut-off is removed our computation can be used to determine
the cut-off dependence of the coupling constant $f$ as $C_n$ are kept
constants.
We find that the beta function obtained in this manner contains the parameters
$C_n$,
\eqn\betaelo{\Lambda{\partial\over\partial\Lambda}f=\beta_0(C_2,C_4)f^2
+O(f^3),}
with
\eqn\betafr{\beta_0(C_2,C_4)=-{1\over2\pi}{1\over{1-{2C_2^2\over C_4}}}.}
It is natural that the presence of the additional coupling constants in
the theory modify the
beta function. What is remarkable in this result is that the dependence
on $C_n$ remains unsuppressed
over all the range of the perturbation expansion, $\mu>\Lambda_\sigma$.
Any dependence on $C_n$ should appear through the combination \insv\ according
to the perturbation expansion on flat or slowly varying background field. But
the beta function contains such dimensionless ratio of the coupling constants
in which the suppression factor of \insv\ is cancelled.
Thus the `irrelevant' coupling constants $C_n$ are not at all irrelevant
when instantons are present.

Additional non-perturbative effects are expected to show up for $C_n=O(1)$.
Since the lagrangian contains new, non positive definite terms the self-dual
saddle points are not necessarily stable. If they
become unstable for certain choice of $C_n$ then we find a phase transition
at those values of the coupling constants. Even if the saddle points
remain stable the mini-instanton action can be negative. For such a choice of
the $C_n$ the mini-instantons condense and the saddle point expansion ceases to
be reliable. The theory becomes similar to frustrated spin models and
the weak coupling expansion is rendered useless despite the presence of
the small $f$.

The organization of the paper is the following. In Section II. the
tree level approximation to the path integral is discussed. The
one-loop quantum fluctuations are considered in Section III. Section
IV. contains the derivation of the beta function for the unit winding
number sector. The renormalization group in the dilute instanton gas
approximation is considered qualitatively in
Section V. Finally our conclusion is presented in Section VI.
\bigskip
\centerline{\bf II. TREE LEVEL}
\medskip
\nobreak
\xdef\secsym{2.}\global\meqno = 1
\medskip
The saddle point approximation for the path integral
\eqn\pist{  \int D[\vec\phi]~\prod_{x}
\delta(\vec\phi^2(x)-1)e^{-S[\vec\phi]},}
starts with the selection of the extrema of the action $S[\phi]$,
on the tree level.
In order to find the saddle point of the action \exta\ which
satisfies the constraint \constr\ we introduce the Lagrange
multiplier field $\lambda(x)$ and use the action
\eqn\alag{S^{\lambda}[\vec\phi] = {1\over2f}\int d^2 x~\Biggl[\Bigl
( 1- {3\over 4}~{C_2\over{\Lambda^2}}
({\partial_\mu\vec\phi})^2
+ {5\over 64}~{C_4\over{\Lambda^4}}
({\partial_\mu \vec\phi})^4
\Bigr )
({\partial_\mu \vec\phi })^2+
\lambda (x)(\vec\phi^{2} - 1) \Biggr ],}
at the tree level computation. The equation of motion,
${{\delta S^{(\lambda)}[\vec\phi]\over\delta\vec\phi(x)} = 0}$,
for the constrained system is then
\eqn\mot{\Bigl ( 1- {3\over 2}~{C_2\over{\Lambda^2}}
({\partial_\mu\vec\phi\partial_\mu\vec\phi})
+{15\over 64}~{ C_4\over{\Lambda^4}}
({\partial_\mu \vec\phi \partial_\mu \vec\phi})^2
\Bigr ) \partial_\mu \partial_\mu \vec\phi
+ 2\lambda(x) f \vec\phi = 0.}
The Lagrange multiplier field is eliminated by the help of \constr,
\eqn\la{\lambda(x) = {- 1\over 2f}
\Bigl ( 1- {3\over 2}~{C_2\over{\Lambda^2}}
({\partial_\mu\vec\phi\partial_\mu\vec\phi})
+{15\over 64}~{ C_4\over{\Lambda^4}}
({\partial_\mu \vec\phi \partial_\mu \vec\phi})^2
\Bigr ) \partial_\mu \vec\phi \partial_\mu \vec\phi,}
and the equation of motion is written in the form
\eqn\emot{\Bigl ( 1- {3\over 2}~{C_2\over{\Lambda^2}}
({\partial_\mu\vec\phi\partial_\mu\vec\phi})
+{15\over 64}~{ C_4\over{\Lambda^4}}
({\partial_\mu \vec\phi \partial_\mu \vec\phi})^2
\Bigr ) \Bigl (- {\partial_\mu \partial_\mu} -
{\partial_\mu\vec\phi\partial_\mu\vec\phi} \Bigr) \vec\phi = 0.}
Notice that due to the particular choice of the irrelevant
operators in \exta, one can factorize out the equation of motion of the
usual lagrangian, \sakez\ in \emot.
Then the one-instanton solution to \emot\ is the well known
Belavin-Polyakov solution \ref\belpolr\ which can be
conveniently parametrized in the following way,
\eqn\sol{\cases{\eqalign{
\phi^{(1)}_{0}={2\rho(x-x_{0})\over\rho^{2} + |z-z_{0}|^{2}}\cr
\phi^{(2)}_{0}={2\rho(y-y_{0})\over\rho^{2} + |z-z_{0}|^{2}}\cr
\phi^{(3)}_{0}={|z-z_{0}|^{2}-\rho^{2}\over\rho^{2} + |z-z_{0}|^{2}}.\cr}}}
Here ($x_{0}$,$y_{0}$) are the center of the instanton coordinates,
$\rho$ is the instanton size. Actually the solution contains six arbitrary
parameters related to the symmetry of the action : two for the translational
invariance, one for the dilatational invariance and three related to rotations
in the internal space. In \sol\ we have displayed only three of them (
translational and dilatational invariance). The fourth parameter, which is
related to the
rotation around the third axis in internal space (which we orient for
convenience in the direction orthogonal to the two dimensional space-time)
and mixes the components $\phi^{(1)}_{0}$ and $\phi^{(2)}_{0}$ is set to
zero. The other two parameters are unimportant for the ratio \irat.
This is because the other two
symmetries of the action in the one-instanton sector are also symmetries
in the homogeneous background sector. Thus the contribution of these
zero modes to $Z_1$ is the same as the contribution to $Z_0$ and they drop out
in \irat.

Inserting
\eqn\lagr{ \partial_\mu\vec\phi_0\partial_\mu\vec\phi_0=
{ 8\rho^{2}\over {(\rho^{2} + |z-z_{0}|^{2})^2}},}
into \exta\ we find the action for the one-instanton configuration,
\eqn\szer{ S[\vec\phi_{0}]\equiv S_{0}(\rho) =
{4\pi\over f} \bigl ( 1 -
{2C_{2}\over\rho^2\Lambda^2}
+{C_{4}\over\rho^4\Lambda^4} \bigr ).}
This result shows the scale dependence arising due to the irrelevant
operators. Why should one be concerned for the impact of irrelevant operators
on the theory ? Quantum Field Theories are plagued by ultraviolet
divergences whose regularization is achieved by the help of irrelevant
operators (c.f. lattice or Pauli-Villars regularization). Thus we always have
some irrelevant operator present in the bare action. Though the
irrelevant operators introduced in this paper do not regulate
the theory the study of their influence on the physical content of
the model will be instructive to appreciate the importance of the regulators
in general.

The irrelevant operators lift the degeneracy of the tree level action
with respect of the dilatation transformation and the only stable
instanton has the size,
\eqn\min{{\overline\rho}^{2} = {C_4\over C_2}{1\over\Lambda^2},}
and action
\eqn\smin {S_{0}(\overline\rho) = {4\pi\over f} \bigl ( 1 - {2C_2^2\over C_4}
\bigr).}
The very existence of this minimum is due to the choice of the values of
$C_n$. If all the coefficients
had the same sign then there would not be any stable instanton
solution at all.
\vskip 15 pt
\par\noindent
\nobreak
\medskip
\centerline{\bf III. ONE LOOP LEVEL}
\xdef\secsym{3.}\global\meqno = 3
\medskip
After identifying the saddle point at the tree level we can turn to the
contribution of the quantum fluctuations in the path integral
\eqn\partf{Z=\int D[\vec\phi]~\prod_{x}
\delta(\vec\phi^{2}(x)-1)e^{-S[\vec\phi]}.}
As long as $C_n \leq O(1)$ we can use the semiclassical approximation and
expand the field $\vec\phi$ around the configuration $\vec\phi_{0}$
\eqn\split{\vec\phi=\vec\phi_{0}+\vec\eta,}
and keep terms at most quadratic in the fluctuation $\vec\eta$.
Inserting \split\ in \partf\ we obtain
\eqn\partsem{\eqalign{ &Z_1 =
\int D [\vec\eta~]~\prod_{x}\delta(\vec\eta(x)\cdot\vec\phi_{0}(x))
{}~~e^{-S_0(\rho)}\cr
&{\rm exp}\Biggl\{- {1\over 2f}
\int d^2x~\vec\eta\Bigl ( 1- {3\over 2}~{C_2\over\Lambda^2}
({\partial_\mu\vec\phi_0})^2
+{15\over 64}~{C_4\over\Lambda^4} ({\partial_\mu \vec\phi_0})^4
\Bigr ) \Bigl (- {\partial_\mu \partial_\mu} -
({\partial_\mu\vec\phi_0})^2 \Bigr)\vec\eta\Biggr\}.\cr}}
Note that the classical term ${\rm exp}\bigl (-S_{0}(\rho) \bigr )$
appears under the integral over the fluctuation $D[\vec\eta]$
and does not factorizes.
In fact, in contrast to the case ~$C_n=0$~, the classical action,
{}~$S_{0}(\rho)$~ depends on the instanton size and the integration
measure $D[\vec\eta]$ contains an integration over the
"collective coordinate" $\rho$.

Eq. \partsem , as it stands, is divergent and needs regularization.
It will be regulated by an appropriate truncation
of the functional space where the fluctuation determinant is calculated.
Had we included a true non-perturbative regulator in the lagrangian
it would have generated finite one-loop contribution around the
instantons with the preferred size. But now we have to be careful and
to employ a regulator for the quantum fluctuations which has $\Lambda$,
the only scale parameter in \partsem\ as the cut-off parameter.

First we factorize the quantum fluctuation operator as follows.
\partsem\ can be written in a condensed form as
\eqn\partsemc{Z_1=e^{-S[\vec\phi_{0}]}
\int [D\vec\eta~]~\prod_{x}\delta(\vec\eta(x)\cdot\vec\phi_{0}(x))
e^{-\vec\eta AB\vec\eta},}
with
\eqn\aop{A(x,y)=\delta(x-y)\Bigl(1-{3\over2}~{C_2\over\Lambda^2}
({\partial_\mu\vec\phi_0}(x))^2+{15\over 64}~{C_4\over\Lambda^4}
({\partial_\mu\vec\phi_0}(x))^4\Bigr),}
and
\eqn\bop{B(x,y)=-{1\over2f}\delta(x-y)\Biggl({\partial_\mu \partial_\mu}-
({\partial_\mu\vec\phi_0}(x))^2\Biggr).}
By ignoring the the zero modes first we write \partsemc\ as
\eqn\partsemd{\eqalign{Z_1&=e^{-S[\vec\phi_{0}]}{\rm det}^{-{1\over2}}A
{\rm det}^{-{1\over2}}B\cr
&=e^{-S[\vec\phi_{0}]}{\rm det}^{-{1\over2}}A
\int [D\vec\eta~]~\prod_{x}\delta(\vec\eta(x)\cdot\vec\phi_{0}(x))
e^{-\vec\eta B\vec\eta},\cr}}
where the determinants are taken in the functional space specified by
the constraint in \partsemc.

The manipulation of the path integral in \partsemd\ is only formal. In fact,
the presence of a negative eigenvalue of $AB$, the indication of the
instability of the saddle points, may not be recognizable by computing
${\rm det}A{\rm det}B$. Furthermore the zero modes arising from the
breakdown of rotations and translations make this separation ambiguous, too.
In order to avoid these problems we impose an additional condition on
the coupling constants of the irrelevant operators,
\eqn\acirop{C_n<<1,}
which makes all eigenvalues of the operator $A$ positive. Thus
the zero modes of $B$ and $AB$ are identical and the usual collective
coordinate method to isolate the zero modes can be applied in \partsem\ and
\partsemd\ in the same manner. Note that the splitting of the degeneracy
of the tree level action under the dilatation transformation does not
necessarily imply that the construction of the collective coordinate
corresponding to the scale of the instanton should
be skipped in the computation.
In our case the minimum of the tree level action as the function
of the instanton size is of fourth order and the one-loop integral
is still ill-defined without keeping the size of the instanton fixed.
Naturally, another bonus of keeping the instanton size as a collective
coordinate is that the comparison with the usual case, $C_n=0$ is
more transparent.

The remaining path integral in the last equation of \partsemd\ can be
evaluated by exploiting the scale invariance of the $\vec\eta B\vec\eta$.
Following Ref. \ref\jevickir\ we write it as
\eqn\oldop{\int d^2x~
\Bigl({1\over2}\partial_\mu\vec\phi_{0}\Bigr)^2
\vec\eta\Bigl(-{{\partial_\mu\partial_\mu}\over{
{1\over2}(\partial_\mu\vec\phi_{0})^2}}-2\Bigr)\vec\eta,}
and we include the first term of the quantum fluctuation operator
in the measure,
\eqn\meas{d\mu(x)=d^2x
\Bigl({1\over2}\partial_\mu\vec\phi_{0}\partial_\mu\vec\phi_{0}\Bigr).}
It is then sufficient to consider the eigenvalue problem for the operator
\eqn\newop{-{{\partial_\mu\partial_\mu}
\over{{1\over2}\partial_\mu\vec\phi_{0}\partial_\mu\vec\phi_{0}}}-2.}
The formal advantage of this separation is that \newop\ is actually the
Laplacian on the sphere. The eigenvalues $E$ in
\eqn\eig{\Bigl (- {{\partial_\mu \partial_\mu}\over{
{1 \over 2}\partial_\mu \vec\phi_{0} \partial_\mu \vec\phi_{0}}
} -2 \Bigr )\vec\psi=E\vec\psi,}
where the eigenmodes $\vec\psi(x)$ are subject to the constraint
\eqn\cossy {\vec\psi(x)\cdot\vec\phi_{0}(x) = 0,}
are known to be \ref\polyjeptr\
\eqn\eigen {E_{j}=j(j+1) - 2,}
with degeneracy
\eqn\deg {2(2j + 1).}

There is a simple geometrical interpretation of the construction above.
The integration measure \meas\ is the Jacobian of the stereographical
projection of the two dimensional space-time onto a sphere with radius
$\rho$. The possibility of this mapping onto a
sphere is the immediate consequence of the conformal invariance of the
theory with $C_n=0$. The eigenvectors of \eig\ form a complete
orthogonal basis on the sphere.

Now we regulate the contribution of the quantum fluctuations. A natural
regularization is that one truncates the spectrum of the
small fluctuations operator in \partsem\ at
\eqn\ineqc{{E_j\over\rho^2}\leq\Lambda^2.}
The factor $\rho^2$ in the denominator restores the original dimension
of the eigenvalues. \ineqc\ represents the omission of
short distance scale modes from the theory with large wave number
on the sphere. The spectrum of the ultralocal operator, $A$ is
finite in coordinate space and one has
\eqn\spau{{\rm lndet}A
= (2\pi\Lambda)^2\int d^2 x~\ln~\Biggl
( 1- {3\over 2}~{C_2\over{\Lambda^2}}(\partial_\mu\phi^a)^2
+ {15\over 64}~{C_4\over{\Lambda^4}}(\partial_\mu\phi^a)^2
(\partial_\mu\phi^a)^2\Biggr)}
The proper evaluation of ${\rm lndet}A$ can only be done in such
regularizations
which are local in space. The measure of the space-time integration is
then made dimensionless by normalizing by the elementary volume corresponding
to a single degree of freedom. In lattice regularization this factor is
$a^2$, the square of the lattice spacing. The factor $(2\pi\Lambda)^2$
is to perform this normalization in terms of our cut-off with dimension
of energy, $\Lambda$. The integral in \spau\ is computed in the Appendix.
It is indeed finite and yields
\eqn\spaf{\eqalign{ {\rm lndet}A =&
\pi(2\pi)^2 \rho\Lambda\times \cr
&\times\Biggl \{ \rho\Lambda\ln~\Biggl [ {(\rho\Lambda)^4\over
(\rho^2\Lambda^2 + \sqrt {15 C_4})^2
- 4\rho^2\Lambda^2\sqrt {15 C_4}
{\rm cos}\biggl ({1\over 2}
\sqrt {{15 C_4\over 36 C_2^2} - 1}\biggr )}\Biggr ]\cr
&+ (15 C_4)^{1\over 4} {\rm cos}\biggl ({1\over 2}
\sqrt {{15 C_4\over 36 C_2^2} - 1}\biggr )\times\cr
&\times\ln~ \Biggl [
{{\rho^2\Lambda^2 - 2\rho\Lambda (15 C_4)^{1\over 4}
{\rm cos}\biggl ({1\over 2}
\sqrt {{15 C_4\over 36 C_2^2} - 1}\biggr ) + \sqrt {15 C_4}}\over
{\rho^2\Lambda^2 + 2\rho\Lambda (15 C_4)^{1\over 4}
{\rm cos}\biggl ({1\over 2}
\sqrt {{15 C_4\over 36 C_2^2} - 1}\biggr ) + \sqrt {15 C_4}}}\Biggr ]\cr
&+2\pi -2{\rm arctg}\Biggl [
{{\rho\Lambda - (15 C_4)^{1\over 4}{\rm cos}\biggl ({1\over 2}
\sqrt {{15 C_4\over 36 C_2^2} - 1}\biggr ) }\over
{(15 C_4)^{1\over 4}{\rm sin}\biggl ({1\over 2}
\sqrt {{15 C_4\over 36 C_2^2} - 1}\biggr )}}\Biggr ]\cr
&-2{\rm arctg}\Biggl [
{{\rho\Lambda + (15 C_4)^{1\over 4}{\rm cos}\biggl ({1\over 2}
\sqrt {{15 C_4\over 36 C_2^2} - 1}\biggr ) }\over
{(15 C_4)^{1\over 4}{\rm sin}\biggl ({1\over 2}
\sqrt {{15 C_4\over 36 C_2^2} - 1}\biggr )}}\Biggr ]\Biggr \}.\cr}}

The regularization of the determinant of the other operator, $B$,
is achieved by keeping only the eigenvalues
\eqn\numcut{j(j+1)-2\leq\rho^2\Lambda^2,}
in the determinant which means the truncation of the functional space in
\partsemd\ by keeping the spherical harmonics up to
\eqn\jmax {j_{max}= [\rho^2\Lambda^2],}
in the expansion of the eigenmodes. Note that $j_{max}$, the integer part
of $\rho^2\Lambda^2$ is a discontinuous function of the cut-off, $\Lambda$.
Since the eigenvalues of $A$ are
$1+O({1\over\rho^2\Lambda^2})$, \numcut\ implies the truncation of the
spectrum of $AB$ at $O(\Lambda^2)$. The appearance of the instanton
size parameter, $\rho$, in these expressions is similar to the usual
dimensional transmutation, observed in the perturbation expansion
around the flat background field. In fact, in both cases it is the
appearance of the cut-off which allows nontrivial scale dependence
in terms of physical scales.

Our cut-off procedure \numcut\ removes the fluctuations with
short wavelength fluctuations on the sphere. Notice that the
ratio of the corresponding distances on the sphere and in the
space-time depends on the locations. In fact, their ratio is unity
only at the contact point between the plane and the sphere.
Imagine a mesh of lattice points on the sphere corresponding to the
lattice regularization, with uniform area, $a^2$. The images of these
points in the space-time form a nonuniform lattice, where the inverse
of the `lattice spacing' is given by
\eqn\sphr{\Lambda(z)=\Lambda{\rho^2\over|z|^2+\rho^2}.}
The drop of the cut-off $\Lambda(z)$ for $|z|>>\rho$ reflects the
growing of the `lattice spacing' with the distance $|z|$ from the center of the
instanton. Thus our ultraviolet regulator tends to suppress the phase space
for the modes which have longer length scale than the instanton size
and serves as an infrared cut-off, too. The translation of the result obtained
by this inhomogeneous cut-off into the conventional computational scheme
based on homogeneous cut-off is not too difficult, it only brings the
multiplicative factor $e^2$ to $Z_1/Z_0$, \ref\forsterr. This finite
correction factor represents the contributions of the long range modes
to the fluctuation determinant \ref\polyplr\ since the homogeneous
cut-off does not suppresses the infrared regime.

We ultimately need the ratio $Z_1/Z_0$ and the infrared cut-off should
not be important there.
Indeed the instanton should represent small perturbation for the modes
which are longer than the instanton size
and the corresponding eigenvalues practically drop from $Z_1/Z_0$.

Finally we arrive to the result,
\eqn\partin{\eqalign{{ Z_1\over Z_0}=~&(2\pi)({16\over3})^2{1\over f^2}
\int_{0}^{\infty}{d\rho\over\rho^3}e^{
-{4\pi\over f} \bigl (1-{2C_2\over {\rho^2\Lambda^2}}+
{C_4\over {\rho^4\Lambda^4}}\bigr)}\cr
&\Biggl [{ \prod_{j>0}\Bigl [ j(j+1)\Bigr ]^{(2j+1)}\over
\prod_{j>1}\Bigl [ j(j+1) - 2\Bigr ]^{(2j+1)}}\Biggr ]~
{\rm det}^{-{1\over2}}A\cr}}
where the constant factors in the first line come from the zero-modes and
the second line contains the fluctuation determinant ratio
\eqn\det{ D(\rho\Lambda)\equiv {\prod_{j>0}\Bigl [ j(j+1)\Bigr ]^{(2j+1)}\over
{\prod_{j>1}\Bigl [ j(j+1) - 2\Bigr ]^{(2j+1)}}},}
as well as the contribution of the ultralocal operator
\eqn\secnd{ {C(\rho\Lambda)\equiv {\rm det}^{-{1\over2}}A }.}

The last integration over the instanton size is carried out in the
saddle point approximation,
\eqn\ro {\int_{0}^{\infty}{d\rho\over\rho^3}
e^{-{4\pi\over f} \bigl (1-{2C_2\over {\rho^2\Lambda^2}}+
{C_4\over {\rho^4\Lambda^4}}\bigr )}D(\rho\Lambda)C(\rho\Lambda).}
It is useful to change integration variable from $\rho$
to $1\over\rho^2$,
\eqn\ros{ e^{-{4\pi\over f} \bigl ( 1 - {2C_2^2\over C_4}
\bigr )}\int_{0}^{\infty} d({1\over\rho^2})
e^{-{4\pi\over f}{C_4\over\Lambda^4}
\bigl ({1\over\rho^2} - {1\over\overline\rho^2}\bigr )^2}
D(\rho\Lambda)C(\rho\Lambda)}

We find here a competition between the exponential gaussian
classical factor and the one-loop determinant
$D(\rho\Lambda)C(\rho\Lambda)$ in the integral,
the latter being a discontinuous function of its argument. This
discontinuity is naturally an artifact of our cut-off procedure,
\numcut. Had we used a consistent regulator which appear in a
uniform manner on the tree and the one loop level, like the lattice
spacing, we would have had smooth dependence on the cut-off. By the
dimensional transmutation this would have given smooth scale dependence,
too. But we believe that this slight inconsistency of our calculation
will not be important from the point of view of our result.
It can be shown \forsterr that for very large values of the
instanton size,
\eqn\lim{\rho^2\Lambda^2 >> 1 }
the determinant becomes the usual smooth function,
\eqn\detlim {D(\rho\Lambda)=e^{\ln\rho^2\Lambda^2}.}
For smaller values of $\rho\Lambda$ $D(\rho\Lambda)$ is a step function.
But in the region $\rho\Lambda<1$ where the cut-offs both in the
tree level and the quantum fluctuations play role the tree level
exponential factor is small. Thus the possible differences
of the way the cut-off appears at these level are suppressed
as long as the typical instanton size is larger than the cut-off,
\eqn\inlcu{C_4>C_2.}

We proceed now to evaluate the contribution to $Z_1/Z_0$ in this region by
performing the gaussian integration in $\rho$. The result is
\eqn\sad{ {1\over4f^2} D(\overline\rho\Lambda)C(\overline\rho\Lambda)
e^{-{4\pi\over f} \bigl ( 1 - {2C_2^2\over C_4}
\bigr )} \Lambda^2 \sqrt {f\over C_4}.}
Note the presence of an extra multiplicative factor $\sqrt{f}$
compared to the case $C_n=0$, due to the scale dependence of the
tree level action.

The ratio $Z_1/Z_0$ obtained in \sad\ play crucial role in the
dilute instanton gas approximation. It refers to the whole system
so its characteristic momentum scale is $\mu=0$. Its dependence
on the coupling constants $C_n$ indicates
that the instanton gas generates new relevant coupling constants,
as was announced in the Introduction.
\nobreak
\medskip
\centerline{\bf IV. BETA FUNCTION}
\xdef\secsym{4.}\global\meqno = 1
\medskip
The standard result for the evolution of $f$ in the cut-off
can immediately be recovered for $C_n=0$
by considering only very large values for the
instanton size, \lim\ in \ros. Large instantons are locally flat inside
and the renormalization is universal. The evolution of the coupling constant
can be found for any large enough instantons, by inspecting the integrand of
\ros\ instead the integral itself. In fact,
the integrand in \ros\ reduces to
\eqn\stand{{\rm exp}\Biggl\{-{4\pi\over f} + \ln\rho^2\Lambda^2 \Biggr\},}
for large instantons. The independence of \stand\ from the cut-off gives
\eqn\flow{{1\over f(\rho)}={1\over f(\Lambda)}
-{1\over2\pi}\ln{\rho\Lambda},}
the well known result about the asymptotic freedom of the theory \polyplr.

The derivation of \stand\ and \flow\ in the one-instanton sector used
two essential assumptions. The first one is that the instanton is large enough
for \lim\ be valid. The second one is that the instanton is small
enough to keep the dominance of the tree level contribution over the
one-loop one. In fact, the infrared sector of the dimensionally
transmuted theory, i.e. with mass scale generated dynamically by the
evolution equation,
\eqn\lamsig{\Lambda_{\sigma}^2=\Lambda ^{2}e^{-4\pi/f(\Lambda)},}
is highly non-perturbative. In order to avoid
the contribution of these non-perturbative modes in $Z_1/Z_0$ we have
to constrain ourselves for not too large instantons.
Thus the evolution \flow\ is reliable only in the region
\eqn\val{{1\over\Lambda}~~  << \rho << ~~{1\over\Lambda_{\sigma}}}

We have three coupling constants in our model, $f$, $C_2$ and $C_4$, their
dependence on the cut-off is determined by imposing renormalization
conditions. In the homogeneous background field sector $C_n$ may be kept
constants at any finite order of the loop
expansion. The standard renormalization procedure then yields
\eqn\posflow{\cases{\eqalign{
&C_1 = constant\cr
&C_2 = constant\cr
&f = f(\Lambda),\cr}}}
where $f(\Lambda)$ is taken from \flow.

On an instanton background the situation is different. The classical saddle
point, a mini-instanton has a well determined size $\bar\rho$,
proportional to the cut-off.
The large instantons are `almost' stable since their action
depends weakly on the size parameter. We integrate up the contributions
of the instantons with different size in the collective coordinate method.
But the action of the mini-instanton
which is proportional to bare value of $1\over f$
is much lower than those of the large instantons as long as the theory is
asymptotically free. Thus $Z_1$ is actually
dominated by the mini-instanton. Since
$Z_1/Z_0$ is proportional to the ratio of the eigenvalues of the operator
$AB$, with characteristic length
on the order of magnitude of the cut-off there is no reason to expect
`universality', i.e. the independence of $Z_1/Z_0$ on the non-renormalizable
coupling constants.

The evolution of the coupling constant $f$ can be determined by the
cut-off independence of the the one-instanton contribution,
\eqn\regr{\Lambda{d\over {d\Lambda}} \Bigg\{{1\over f^2}
e^{-{4\pi\over f} \bigl ( 1 - {2C_2^2\over C_4}\bigr )} \Lambda^2
\sqrt {f\over C_4 }\Bigg\} = 0 }
In order to determine the renormalization of $C_n$ we
would need additional renormalization conditions. We shall be contented
in this paper by demonstrating the relevance of the non-renormalizable
coupling constants in general, so no additional renormalization
conditions will be imposed and the values of $C_n$ will be kept as
parameters in investigating the evolution equation for $f$.

In the absence of a general method to decide the relevance of an operator
on should rely on the general blocking procedure in renormalizing
the theory. The trajectory we shall follow, \posflow\
is one of infinitely many possible trajectories in the coupling constant
space and certainly not the result of any blocking procedure. Nevertheless
we shall find a dependence of the trajectory on the initial values of $C_n$,
indicating
the non-negligible effects the non-renormalizable operators play in the
theory. For example, by inspecting \regr\ one finds that the flow

\eqn\othflow{\cases{\eqalign{
&C_2 \propto (\ln\Lambda)^{-1}\cr
&C_4 \propto (\ln\Lambda)^{-1}\cr
&f \propto (\ln\Lambda)^{-1},\cr}}}
is actually compatible with the standard scaling, \lamsig. What is
nevertheless important is that one actually needs a particular fine-tuning
of the non-renormalizable coupling constants in order to maintain the
independence of the one-instanton contribution on the cut-off in the
region \val.

The renormalization condition \regr\ gives the beta function for the
flow \posflow,
\eqn\betaf{\Lambda{d~f\over{d\Lambda}} = -{f^2\over {2\pi}}{1\over
{1-{2C_2^2\over C_4} - {3f\over 8\pi}}}.}
In leading order in $f$ we can neglect the $f$ dependence of the denominator.
Our final expression for the beta function is
\eqn\betafu{\beta(f)=-{f^2\over2\pi}{1\over{1-{2C_2^2\over C_4}}}.}
Due to the conditions \acirop\ and \inlcu\ the non-renormalizable
operators influences only the magnitude of the leading term of the
beta function without flipping its sign. The beta function, obtained
by the saddle point approximation in the instanton scale integration is
independent of the value of the fluctuation determinant. Thus \betafu\ is
actually valid for a wider range of scales than the standard result \flow\
for $C_n=0$, c.f. \val,
\eqn\valm{\rho << ~~{1\over\Lambda_\sigma}.}
\nobreak
\medskip
\centerline{\bf V. INSTANTON GAS}
\xdef\secsym{5.}\global\meqno = 1
\medskip
It is instructive to go beyond the computation of $Z_1/Z_0$
and to consider the dilute
instanton gas approximation. The instanton gas is in principle
as infrared unstable in the presence of the non-renormalizable
operators as in their absence. It appears to us more appropriate to
follow the blocking procedure in the saddle point approximation.
In the blocking one integrates out the variables within a block
with the constraint that the block field has a given value.
This integration can be performed in the semiclassical approximation.
Such a procedure eliminates the instantons which are completely
inside of a block. When only
modes with finite distance scales are eliminated in the saddle point
approximation then the infrared singularities are naturally absent from the
renormalization group equation. In this way one has the hope to generate
the effective coupling constants which are necessary for the stabilization
of the infrared sector by the help of the blocking procedure.

The contribution of a mini-instanton to
the blocking equation when the cut-off is lowered, $\Lambda\to\Lambda'$
can in principle be found in a simple manner if the instanton size $\rho$
satisfies
\eqn\insineq{\Lambda^{-1}<<\rho<<\Lambda'^{-1}.}
The lower bound for the instanton size comes from the requirement that
the original, finer lattice should resolve the instanton. The upper
limit is to have an instanton which is eliminated completely by the
blocking. When this inequality is respected then the instanton field is
negligible outside of the new block and the effects of the instanton can fully
be incorporated by the appropriate choice of the new coupling constants.
In fact, if the instanton field is non-negligible on the blocked configuration
then the separation of the modes to be eliminated from the ones to be kept
is problematical and leads to a more complicated blocked action.

The theory with appropriately chosen $C_2,~C_4>0$ is better suited to the
blocking as the one with $C_n=0$. This is due to the clear separation
of the mini-instanton peak of the tree level exponential factor in \ros\
and the infrared region where the quantum fluctuations become dominant.
In the integration over the instanton size in \ros\ we have two large
contribution.  One comes from the mini-instanton peak at
$\lambda=(\rho\Lambda)^{-2}=(\bar\rho\Lambda)^{-2}$. The width of this peak is
$\Delta\lambda=O(\sqrt{f})$. Another region where the integrand is large
is in the infrared, $\lambda=O((\Lambda_\sigma/\Lambda)^2)$ where
the fluctuation determinant starts to grow. These two regions are
separated by the distance $\lambda=O(1)$. Thus there is a conveniently
wide region in $\lambda$ which separates the mini-instanton
from the dangerous infrared region. The elimination of
variables in the blocking $\Lambda\to\Lambda'$ with
\eqn\blsre{{\Lambda_\sigma\over\Lambda}<{\Lambda'\over\Lambda}<1,}
then can safely be performed in the semiclassical approximation.

The instanton gas is dominated by the zero winding number sector,
by an equal density of instantons and anti-instantons. Each of
those are mini-instanton and generate the beta function discussed in the
previous section. Thus the renormalization of the complete theory
is characterized by the renormalization group equations which are
sensitive to the non-renormalizable coupling constants, $C_n$. The truly
flat background field case which supports the $C_n$ independent beta function
is unimportant due to the entropy of the instantons.

It is interesting to speculate what happens as the non-renormalizable
coupling constants are increased to $C_n=O(1)$. The small fluctuations
around the instanton background may become unstable in this case.
In fact, for such values of the coupling
constants certain elements of the diagonal matrix \aop\ might be negative.
The destabilization of the mini-instantons triggers a phase transition,
i.e. non-analytic dependence of the infrared quantities on the coupling
constants.

If the instanton gas approximation remains stable as
the values of $C_n$ are increased then one finds another, more
intriguing phenomenon. Depending on
the ratio $C_2^2/C_4$, the instanton action can be rather small and the
tree level exponential factor in \ros\ may give a large fugacity for the
instantons. For $C_4<2C_2^2$ the action of the mini-instanton
is even negative. This drives the condensation of the instantons which
in turn makes the dilute gas approximation definitely invalid. The theory then
becomes similar to a frustrated spin model and the weak coupling expansion
ceases to be useful despite the presence of the small coupling constant
$f$.
\nobreak
\medskip
\centerline{\bf VI. CONCLUSION}
\xdef\secsym{6.}\global\meqno = 1
\medskip
The semiclassical expansion was reconsidered for an asymptotically free
model, the two dimensional sigma model in this paper. There is
only one relevant operator on a homogenous background field,
the usual lagrangian of the model.
But as one looks into the unit winding number sector the instanton
action is found to be dependent on the higher dimensional operators
or the regulator.
This is because the effects of these higher dimensional operators
such as a true non-perturbative regulator appear already on
the tree level, in the bare lagrangian.
In particular the degeneracy of the instanton action due to the
scale invariance of the renormalized action is lifted by the
higher dimensional operators. Thus the instantons might either be completely
destabilized or only be stabilized at a certain scale which can be close to
the ultraviolet cut-off. In the latter case, the saturation of the path
integral by the mini-instantons produces a non-universal dependence
of the non-renormalizable coupling constants.
The beta function of the theory is found to be dependent on
to the choice of the higher dimensional operators.

The through determination of the
renormalization group flow requires the complete blocking or the imposition
of sufficient number of renormalization conditions. We followed neither of
these complete procedures and computed the beta function only for the
asymptotically free coupling constant of the model.
The dependence of this beta function on the non-renormalizable
coupling constants
is not surprising in the ultraviolet regime. What is unexpected in
our result is the absence of the usual power suppression
of the non-renormalizable coupling constants. In fact, note that the
first coefficient of the beta function is determined by the
tree level value of the instanton action. This is a dimensionless
number and is independent of the scale of the non-renormalizable
coupling constants. The
mini-instantons reflect certain ultraviolet effects slightly below the cut-off.
But as the cut-off is lowered the mini-instantons remain present below the
current cut-off and ultimately generate modifications all over the
ultraviolet scaling regime. The lower edge of the asymptotical scaling
regime is cut-off independent so we expect the dependence on the
new coupling constants down into the infrared regime of the theory.
We believe that
the computation of the anomalous dimensions in the
framework of the dilute instanton gas picture would indicate the
relevance of these operators in agreement with our results.
These features are missed by the dimensional regularization and can only be
detected by schemes which allow non-perturbative manipulations
on the path integral.

We included higher dimensional operators in the action which do not
influence the shape of the saddle point. One can easily introduce
other higher derivative terms which do actually deform the saddle
point configuration. The self-duality of the instantons is lost
for such models. One expects then important modification of the
dynamics of light fermions compared to the usual instanton gas description.

In the case of the Yang-Mills theory \hdrf\ these operators
serve as ultraviolet cut-off for higher loop graphs.
This is a demonstration of the fact that
any regulator is after all an `irrelevant' operator at the ultraviolet
fixed point. Our results thus raises some questions to
settle in the future about the equivalence
of different regularization schemes and the sufficiency of considering
only renormalizable field theoretical models.
\goodbreak
\bigskip
\centerline{\bf ACKNOWLEDGEMENT}
\medskip
\nobreak
\par The authors thank Pierre van Baal and Laurent Lellouch for valuable
discussions.
\bigskip
\vfill
\eject
\goodbreak
\centerline{\bf REFERENCES}
\medskip
\nobreak
\item{\gellowr}\gellow
\item{\wilsrgr}\wilsrg
\item{\zinnjr}\zinnj
\item{\chiranr}\chiran
\item{\miransr}\mirans
\item{\bardeenr}\bardeen
\item{\thoofir}\thoofi
\item{\polyir}\polyi
\item{\parisr}\paris
\item{\zinninr}\zinnin
\item{\reiszr}\reisz
\item{\belpolr}\belpol
\item{\jevickir}\jevicki
\item{\polyjeptr}\polyjept
\item{\forsterr}\forster
\item{\polyplr}\polypl
\vfill
\eject
\centerline{\bf APPENDIX}
\xdef\secsym{A.}\global\meqno = 1
\bigskip
\nobreak
In this appendix we present the computation of
{\rm lndet}$A(x,y)$, where the determinant is ment in the functional sense and
A(x,y) is the ultralocal operator given in \aop,
\eqn\tracelog{ {\rm lndet}A(x,y)
= (2\pi\Lambda)^2\int d^2 x~\ln~\Biggl
( 1- {3\over 2}~{C_2\over{\Lambda^2}}(\partial_\mu\phi^a)^2
+ {15\over 64}~{C_4\over{\Lambda^4}}(\partial_\mu\phi^a)^2
(\partial_\mu\phi^a)^2\Biggr).}
As pointed out in Section III, the integral has to be normalized by the
factor $(2\pi\Lambda)^2$. Inserting \lagr\ in \tracelog\ we have
\eqn\tracelo{\eqalign{{\rm lndet}A(x,y)
=& (2\pi\Lambda)^2\int d^2 x\cr
&\ln~\Biggl( 1- 12~{C_2\over{\Lambda^2}}
{\rho^2\over{(\rho^2 + |z-z_0|^2)^2}}+ 15~{C_4\over{\Lambda^4}}
{\rho^4\over{(\rho^2 + |z-z_0|^2)^4}} \Biggr ).\cr}}

By introducing the dimensionless variable
\eqn\varia {y = \Lambda^2(\rho^2 + |z-z_0|^2),}
we can rewrite our integral in the simpler form
(the factor $\pi$ comes from the angular integration)
\eqn\tracel{\eqalign{
{\rm lndet}A(x,y)&=\pi(2\pi)^2 \Lambda\rho\int_{\rho\Lambda}^{\infty}
dy~\ln~\biggl ( 1 - 12 {C_2\over y^2} + 15 {C_4 \over y^4} \biggr )\cr
&=\pi(2\pi)^2 \Lambda\rho
\int_{\rho\Lambda}^{\infty} dy~
\Biggl \{ \ln~\biggl ( y^4 - 12 C_2 y^2 + 15 C_4 \biggr )
-~\ln~y^4 \Biggr \}.\cr}}
The discriminant of the quartic polynomial above is
\eqn\discr{{\Delta\over 4} = 36 C_2^2 - 15 C_4.}

The constants $C_2<C_4$ are small so \discr\ is negative, i.e. the
argument of the logarithm is always positive and we have
four complex roots for the quartic polynomial. By denoting
\eqn\algam{\cases{\eqalign{
\alpha = {(15 C_4)^{1\over 4} {\rm cos}\biggl ({1\over 2}
\sqrt {{15 C_4\over 36 C_2^2} - 1}\biggr ) }\cr
\gamma = {(15 C_4)^{1\over 4} {\rm sin}\biggl ({1\over 2}
\sqrt {{15 C_4\over 36 C_2^2} - 1}\biggr ) },\cr}}}
we can write
\eqn\trace{{\rm lndet}A(x,y) = \pi(2\pi)^2 \Lambda\rho
\int_{\rho\Lambda}^{\infty} dy~ \Biggl \{
\ln~ \bigl [ (y - \alpha)^2 + \gamma ^2 \bigr ]
+ \ln~ \bigl [ (y + \alpha)^2 + \gamma ^2 \bigr ]
- \ln~y^4 \Biggr \}. }

After performing the integral we find
\eqn\trac{\eqalign{{\rm lndet}A(x,y)&=\pi(2\pi)^2 \Lambda\rho\cr
&\Biggl \{
\rho\Lambda \ln~ \Biggl [ {(\rho\Lambda)^4\over
{\bigl [ (\rho\Lambda - \alpha)^2 + \gamma ^2 \bigr ]
\bigl [ (\rho\Lambda - \alpha)^2 + \gamma ^2 \bigr ]}} \Biggr ]
+ \alpha \ln~ \Biggl [ {(\rho\Lambda - \alpha)^2 + \gamma ^2 \over{
(\rho\Lambda + \alpha)^2 + \gamma ^2 }} \Biggr ] +\cr
&+2\pi - 2 {\rm arctg}{{\rho\Lambda - \alpha}\over \gamma }
       - 2 {\rm arctg}{{\rho\Lambda + \alpha}\over \gamma }\Bigg \}.\cr }}

It is easy to verify that in the limit $C_n\to0$
we recover the standard result, ${\rm det}A=1$.
\vfill
\end